# Photon Bell-State Analysis Based on Semiconductor-Superconductor Structures.

Evyatar Sabag, Shlomi Bouscher, Raja Marjieh, and Alex Hayat,

*Department of Electrical Engineering, Technion, Haifa 32000, Israel*

We propose a compact and highly-efficient scheme for complete Bell-state analysis using two-photon absorption in a superconducting proximity region of a semiconductor avalanche photodiode. One-photon transitions to the superconducting Cooper-pair based condensate in the conduction band are forbidden, whereas two-photon transitions are allowed and are strongly enhanced by superconductivity. This Cooper-pair based two-photon absorption results in a strong detection preference of a specified entangled state. Our analysis shows high detection purity of the desired Bell state with negligible false detection probability. The theoretically-demonstrated concept can pave the way towards practical realizations of advanced quantum information schemes.

Entangled states are one of the most counter-intuitive concepts in quantum mechanics which contradict the local realism of classical physics [1,2]. Furthermore, the rapidly developing quantum information science relies on the ability to generate and characterize entangled states [3,4,5,6,7]. The most widely used physical realization of quantum information employs photons as qubits, where the information encoding or entanglement is in polarization [8]. Bell state analysis [9,10] is crucial for characterizing entanglement as well as for quantum information applications based on entanglement, including quantum repeaters and teleportation [11,12,13]. However, it was proven that using linear optics full Bell-state analysis cannot be realized [14], whereas conventional nonlinear optical schemes [15] are significantly less efficient. Superconducting optoelectronics is an emergent field, focused on light-matter interaction in structures combining superconductivity and semiconductors [16,17]. Such combinations were shown to result in strongly enhanced quantum and classical nonlinear optical processes such as spontaneous photon-pair emission [18], enhanced two-photon gain [19] and highly-efficient entangled-photon pair generation [20].

Here we propose a new concept of efficient full Bell-state analysis based on photon-pair detection in a semiconductor structure in proximity to a superconductor. In the proposed scheme, a layer of superconductivity is induced in the semiconductor by the proximity effect [21,22], so that the electrons in the semiconductor are in a Bardeen-Cooper-Schrieffer (BCS) state with a superconducting energy gap at the Fermi level [23]. We show that one-photon absorption is forbidden for photons with energy corresponding to excitation of single-particle states within the superconducting gap. Therefore, at such energies only two-photon absorption into Cooper-pair based BCS state can occur (Fig. 1 a) with rates enhanced by many orders of magnitude compared to other nonlinear processes. Moreover, we show that in a semiconductor quantum-well (QW) in proximity to a superconductor, only a specific circular-polarization-entangled photon-pair state can be absorbed due to total angular momentum conservation, energy conservation and the conduction band (CB) electron spin-entangled states in BCS. Furthermore, we show that this system only detects one specific Bell state $|\Psi^+\rangle$, while being transparent to other Bell states. Energy conservation in two-photon absorption determines the total transition energy, but not the individual photon energies. Therefore, the polarization-entangled photons detected in two-photon absorption can be tagged by different energies, with the corresponding Bell state basis:

$$|\Psi^\pm\rangle = \frac{1}{\sqrt{2}}\left(|R\rangle_{\omega_\mu}|L\rangle_{\omega_\nu} \pm |L\rangle_{\omega_\mu}|R\rangle_{\omega_\nu}\right) \qquad |\Phi^\pm\rangle = \frac{1}{\sqrt{2}}\left(|R\rangle_{\omega_\mu}|R\rangle_{\omega_\nu} \pm |L\rangle_{\omega_\mu}|L\rangle_{\omega_\nu}\right) \qquad (1)$$

In typical direct-bandgap bulk semiconductors, the light-hole (LH) and heavy-hole (HH) valence bands (VB), with angular momentum $J_Z^{LH} = \pm 1/2$ and $J_Z^{HH} = \pm 3/2$, are degenerate [24], allowing the absorption of various two-photon states. However, in a semiconductor QW the LH-HH degeneracy is lifted allowing light-matter interaction only with a specific entangled-photon pair [20], which allows the device to distinguish between $|\Psi^\pm\rangle$ and $|\Phi^\pm\rangle$. Choosing the two-photon energy to match a double excitation from the HH to the superconducting gap allows the absorption of $|\Psi^\pm\rangle$ only, based on energy and total angular momentum conservation alone. Furthermore, we show that the BCS state in the CB allows the absorption of $|\Psi^+\rangle$ only. Our calculation is based on a full quantum optical treatment and a complete BCS model, and our results show strong enhancement of the Bell-state detection efficiency with respect to the false detection events at lower temperatures and for larger LH-HH separation, while taking into account the effects of disorder-induced parasitic one-photon absorption in the superconducting gap.

The detection of the entangled-photon states can be implemented by attaching a superconducting contact to the n-type absorbing region of a standard telecom-wavelength avalanche photodiode (APD) (Fig. 1 b). The rest of the device typically has a wider bandgap to prevent breakdown in the high-field impact ionization avalanche regions (e.g. InP), and thus will not absorb the photons which are absorbed in the narrower-bandgap absorption region (e.g. InGaAs). Absorption of a single photon pair in the n-type region does not affect the CB carrier density. However, the n-type region VB under reverse bias has essentially no hole population. Therefore, a single pair of holes, generated in the absorption of an entangled photon pair, will be accelerated towards the impact-ionization layer and initiate an avalanche resulting in a macroscopic signal. Such an APD, therefore, will selectively detect one specific Bell state, while being transparent to the other three. The other three Bell states can be converted into the detectable state by a simple scheme based on diffraction gratings, two quarter-wave plates (QWP) and a half wave plate (HWP) (Fig. 1 c).

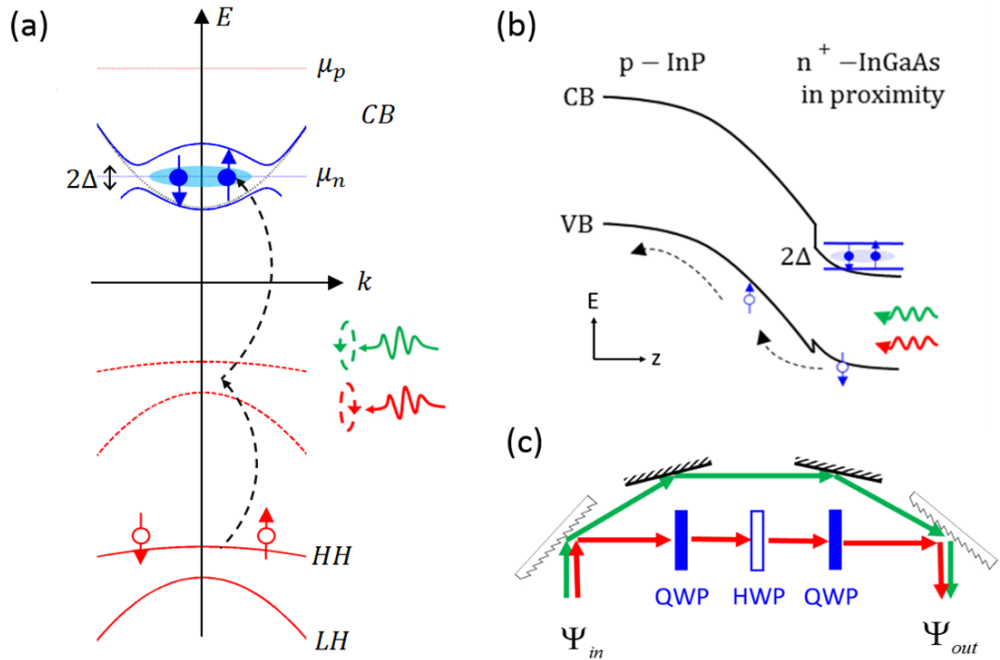

**Figure 1.** (a) Energy band diagram of entangled two-photon absorption in a semiconductor QW superconducting proximity region. (b) Spatial energy band diagram of a standard APD, placed in proximity with a superconductor. (c) An optical scheme converting Bell states into each other.

The two energies are separated on a grating so that only one energy component undergoes polarization manipulation. The QWP transforms circular polarization into linear, and the HWP transforms vertical polarization into horizontal and vice versa when set at 45º. Followed by the second QWP this configuration can transform the Bell states $|\Psi^\pm\rangle$ and $|\Phi^\mp\rangle$ into each other (Eq. 1). Setting the HWP at 0º changes the sign of one term in a Bell state, thus transforming $|\Psi^+\rangle$ and $|\Psi^-\rangle$, as well as $|\Phi^+\rangle$ and $|\Phi^-\rangle$ into each other.

Our theoretical modeling of the APD with proximity-induced superconductivity region is based on a full quantum analysis of two-photon detection. In our model a two-photon state is coupled into the superconductor-induced proximity region in a direct band gap semiconductor [25], the superconducting gap 2Δ is in the semiconductor conduction band in a BCS state, while the valence band is in the normal state of heavy holes and light holes. The hole generation rate in our model using perturbation theory is identical to the photon absorption rate due to the light-matter interaction Hamiltonian (with $\hbar = c = 1$):

$$H_I = \sum_{\boldsymbol{k},\boldsymbol{q},\sigma,J} B_{\boldsymbol{k},\boldsymbol{q}} b_{-(\boldsymbol{k}-\boldsymbol{q}),-J} c_{\boldsymbol{k},J+\sigma} a^\dagger_{\boldsymbol{q},\sigma} + H.c. \qquad (2)$$

where $J$ is the spin-orbit coupled angular momentum, $\sigma$ is the photon circular polarization, $B_{\boldsymbol{k},\boldsymbol{q}}$ the coupling energy, $b^\dagger_{\boldsymbol{k},J}$ and $c^\dagger_{\boldsymbol{k},J}$ are the hole and electron creation operators, respectively, with crystal momentum $\boldsymbol{k}$ and angular momentum $J$, and $a^\dagger_{\boldsymbol{q},\sigma}$ is the photon creation operator with linear momentum $\boldsymbol{q}$ and polarization $\sigma$. The unperturbed Hamiltonian is described by $H_0 = \sum_{\boldsymbol{q},\sigma} \omega_q \left(a^\dagger_{\boldsymbol{q},\sigma} a_{\boldsymbol{q},\sigma} + \frac{1}{2}\right) + \sum_{\boldsymbol{p},J'} \epsilon_{\boldsymbol{p},J'} b^\dagger_{\boldsymbol{p},J'} b_{\boldsymbol{p},J'} + \sum_{\boldsymbol{k},J} \epsilon_k c^\dagger_{\boldsymbol{k},J} c_{\boldsymbol{k},J}$. In order to calculate the rate of hole generation from which we derive the Bell-state detection rate, we use the hole number operator $N_h = \sum_{\boldsymbol{p},J'} b^\dagger_{\boldsymbol{p},J'} b_{\boldsymbol{p},J'}$. The time dependence of the hole number is calculated using $N_h$ expectation value. On the basis of second-order perturbation theory the hole number time-dependent part is $\langle N_h \rangle = \langle N_h(1) \rangle + \langle N_h(2) \rangle$, with $\langle N_h(1) \rangle$ and $\langle N_h(2) \rangle$ the hole number expectation value correction to the first- and second-order in perturbation theory, respectively (for full derivation see [26]),

$$\langle N_h(1) \rangle = \int_{-\infty}^{t} dt_1 \int_{-\infty}^{t} dt_2 \langle \chi_0 | H_I(t_1)[N_h, H_I(t_2)] | \chi_0 \rangle \tag{3}$$

$$\langle N_h(2) \rangle = \int_{-\infty}^{t_2} dt_1 \int_{-\infty}^{t} dt_2 \int_{-\infty}^{t} dt_3 \int_{-\infty}^{t_3} dt_4 \langle \chi_0 | H_I(t_1) H_I(t_2)[N_h, H_I(t_3) H_I(t_4)] | \chi_0 \rangle$$
$$- \int_{-\infty}^{t_2} dt_1 \int_{-\infty}^{t_3} dt_2 \int_{-\infty}^{t} dt_3 \int_{-\infty}^{t} dt_4 \left\langle \chi_0 \left| \begin{array}{l} H_I(t_1) H_I(t_2) H_I(t_3)[N_h, H_I(t_4)] \\ +H_I(t_4)[N_h, H_I(t_3) H_I(t_2) H_I(t_1)] \end{array} \right| \chi_0 \right\rangle \tag{4}$$

where $|\chi_0\rangle = |Ph\rangle|FS\rangle|BCS\rangle$, $|Ph\rangle$ represents the photonic state, (e.g.$|\Psi^\pm\rangle$, $|\Phi^\pm\rangle$), $|FS\rangle$ is the Fermi sea of holes, and $|BCS\rangle$ is the superconducting BCS electron state. The BCS unperturbed Hamiltonian is $H_{BCS} = \sum_{\mathbf{k},J} E_k \gamma_{\mathbf{k},J}^\dagger \gamma_{\mathbf{k},J}$ where $\gamma_{\mathbf{k},J}^\dagger$ is the Bogoliubov quasiparticle creation operator given by the Bogoliubov transformation:

$$c_{\mathbf{k},J}^\dagger(t) = e^{i(E_c + \mu_n)t}\left(u_k e^{iE_k t} \gamma_{\mathbf{k},J}^\dagger - s_J v_k e^{-iE_k t} \gamma_{-\mathbf{k},\bar{J}}\right) \tag{5}$$

where $E_k = \sqrt{\xi_n^2(k) + \Delta^2}$, $\xi_n(k) = \frac{k^2}{2m_n} - \mu_n$, $2\Delta$ is the superconducting gap, $\mu_n$ is the electron quasi Fermi level, $m_n$ is the electron mass, $E_c$ is the edge of the conductance band, $s_J = 1(-1)$ for $J = \uparrow (\downarrow)$, and $u_k(v_k) = \{[1 + (-1)\xi_n(k)/E_k]/2\}^{1/2}$. The calculations in Eqs. (3) and (4) can be described by the one- and two-loop Feynman diagrams corresponding to the first- and second-order perturbation terms (Fig. 2).

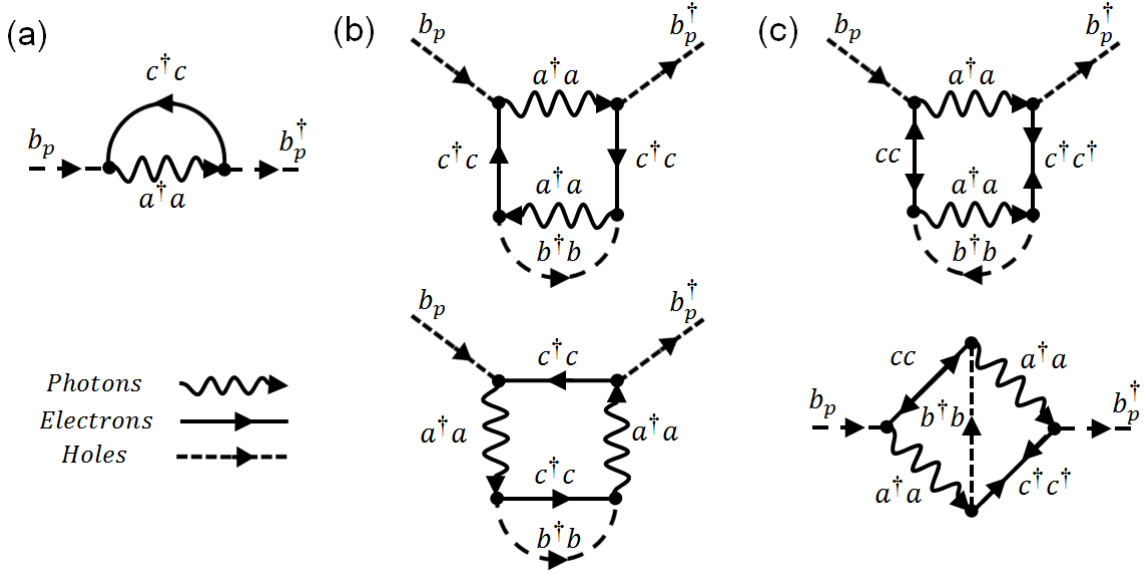

**Figure 2.** Feynman diagrams of the one- and two-loop correction to the hole propagator used to calculate the hole number $N_h$. The solid lines indicate electrons, the dashed lines indicate holes, and the wavy lines indicate photons. (a) The one-loop correction diagram to the hole propagator. (b) The two-loop correction diagrams to the hole propagator without a Cooper pair. (c) The two-loop correction diagrams to the hole propagator involving a Cooper pair.

Starting with the one-photon absorption contribution to the hole generation rate, we obtain that for the low temperatures required for superconductivity ($T \sim 10K$) it practically vanishes for properly chosen energies. However, for broader photon bandwidth this one-photon absorption can result in non-negligible detection rates. All four different Bell-states tagged by two specific energies result in equal contribution from this one-photon process, due to the fact that one-photon processes depend only on the individual photon energies. Calculating the hole generation rate using $R = d\langle N_h \rangle/dt$ and neglecting the hole Fermi-Dirac distribution, due to the low temperatures required for superconductivity ($\sim 10K$) and the extremely negative hole quasi-Fermi level at the N side of the junction ($-\mu_p \sim 1eV$). The resulting one-photon parasitic rate, which is negligible for the appropriate photon energies (for full derivation see [26]),

$$R^{(1)}_{\Psi^\pm,\Phi^\pm} \propto \sum_{J,\xi_{p,J,q_\mu}} |B_{q_\mu}|^2 \, sign(\widetilde{\omega}_{q_\mu} + \mu_J) \Theta(\xi_{p,J,q_\mu} + \mu_p) \Theta(\Lambda_{J,q_\mu})$$
$$\times \left[(1 - f^n_{\xi_{n,J,q_\mu}}) \Theta(\widetilde{\omega}_{q_\mu} - \xi_{p,J,q_\mu}) - f^n_{\xi_{n,J,q_\mu}} \Theta(\xi_{p,J,q_\mu} - \widetilde{\omega}_{q_\mu})\right] + (q_\mu \to q_\nu) \qquad (6)$$

where $\xi^{(\pm)}_{p,J,q} = \frac{1}{1-\overline{m}_J^2}[\widetilde{\omega}_q + \overline{m}_J^2 \mu_J \pm \sqrt{\Lambda_{J,q}}]$, $f^n_{\xi_{n,J,q}} = \left[\exp\left(\beta\sqrt{\xi^2_{n,J,q} + \Delta^2}\right) + 1\right]^{-1}$ is the quasi-particles distribution, with $\xi_{n,J,q} = \overline{m}_J(\xi_{p,J,q} + \mu_J)$ and we've defined $\overline{m}_J \equiv \frac{m_{p,J}}{m_n}$, $\mu_J \equiv \mu_p - \Delta\omega_{p,J} - \frac{m_n}{m_{p,J}}\mu_n$, $\widetilde{\omega}_q \equiv \omega_q - (E_g + \mu_n + \mu_p)$ and $\Lambda_{J,q} \equiv \overline{m}_J^2(\widetilde{\omega}_q + \mu_J)^2 + \Delta^2(1 - \overline{m}_J^2)$, where $m_{p,\pm\frac{1}{2}(\pm\frac{3}{2})}$ is the LH(HH) mass and $\Delta\omega_{p,\pm\frac{3}{2}} = 0, \Delta\omega_{p,\pm\frac{1}{2}} = \Delta\omega_p$ is the energy splitting between the two hole energy bands. Although this contribution appears to yield false detections, a more careful examination of this expression reveals that it vanishes for a large range of energies. Moreover, we show that even with disorder-induced broadening, parasitic one-photon absorption is much weaker than entangled-photon pair absorption – by ~5 orders of magnitude.

Next we consider the superconductivity-enhanced absorption of the photonic state $|Ph\rangle = |\Psi^\pm\rangle$. In our calculation we may neglect terms that do not include Cooper-pair effects, since they describe the same process of one photon absorption, and give a negligible second-order correction to $R^{(1)}_{\Psi^\pm,\Phi^\pm}$. Under this assumption $\langle N_h(2) \rangle$ vanishes for $|\Psi^-\rangle$, but not for $|\Psi^+\rangle$. Calculating the rate of hole generation under the sound assumption that the hole Fermi-Dirac distribution is zero as before, the resulting desired Bell-state detection rate given by hole generation rate (for full derivation see [26]),

$$R^{(2)}_{\Psi^+} \propto \frac{|B_{q_\mu} B_{q_\nu}|^2 \Delta^2 \Theta(\omega_{q_\mu} + \omega_{q_\nu} - 2(E_g + \Delta\omega_p + \mu_n))}{\left(\Delta\omega_{q_\mu,q_\nu} + \Omega^{LH}\right)^2 \left(\Delta\omega_{q_\mu,q_\nu} - \Omega^{LH}\right)^2} + (LH \leftrightarrow HH) \qquad (7)$$

where, $\left(\Omega^{LH(HH)}\right)^2 = \left[\frac{m_p^{LH(HH)}}{m_n}\left(\omega_{q_\mu} + \omega_{q_\nu} - 2(E_g + \Delta\omega_p + \mu_n)\right) - 2\mu_n\right]^2 + 4\Delta^2$ with $m_p^{LH(HH)}$ the LH(HH) mass and $\Delta\omega_{q_\mu,q_\nu} = \omega_{q_\mu} - \omega_{q_\nu}$. This Cooper-pair based hole generation rate is proportional to $\Delta^2$, therefore it vanishes for temperatures higher than the superconducting critical temperature ($T_c$) where $\Delta$ vanishes. Another attribute worth mentioning is the resonance attained by the rate as $\Delta\omega_{q_\mu,q_\nu}$ approaches $\pm\Omega^{LH(HH)}$.

Using the same derivation, both $|\Phi^-\rangle$ and $|\Phi^+\rangle$ result in no contribution for proper photon energies. Practical realizations of entangled photon pairs typically have finite photon bandwidth. If the photon spectrum is too broad, a parasitic absorption can result in a finite false detection contribution from $|\Phi^\pm\rangle$. Under the same assumptions as before the parasitic rate, which practically vanishes for properly chosen energies, is (for full derivation see [26]):

$$R^{(2)}_{\Phi^\pm} \propto \frac{\left|B_{q_\mu}B_{q_\nu}\right|^2 \Delta^2 \Theta\left(\omega_{q_\mu} + \omega_{q_\nu} - 2\left(E_g + \frac{1}{2}\Delta\omega_p + \mu_n\right)\right)}{\left(\Delta\omega_{q_\mu,q_\nu}^{LH} + \Omega\right)^2 \left(\Delta\omega_{q_\mu,q_\nu}^{LH} - \Omega\right)^2} + (LH \leftrightarrow HH) \tag{8}$$

Where $\Delta\omega_{q_\mu,q_\nu}^{LH(HH)} \xrightarrow{m_p^{HH} = m_p^{LH}} \Delta\omega_{q_\mu,q_\nu} \mp \Delta\omega_p$ with the – and + signs for LH and HH, respectively, $\Omega^2 = \left[\frac{m_p}{m_n}\left(\omega_{q_\mu} + \omega_{q_\nu} - 2\left(E_g + \frac{1}{2}\Delta\omega_p + \mu_n\right)\right) - 2\mu_n\right]^2 + 4\Delta^2$ with $2m_p^{-1} = \left(m_p^{LH}\right)^{-1} + \left(m_p^{HH}\right)^{-1}$. Similarly to the $|\Psi^\pm\rangle$ rate, this rate is proportional to $\Delta^2$ and attains a resonance as $\Delta\omega_{q_\mu,q_\nu}^{LH(HH)}$ approaches $\pm\Omega$. The main property of this rate is the requirement for higher photon energy to get a finite term in comparison with $R^{(2)}_{\Psi^+}$ seen by the Heaviside step function, meaning, for properly chosen energies only $|\Psi^+\rangle$ is detected.

Using our results for two-photon detection combined with the one-photon detection rate we define the detection purity (DP) of the $|\Psi^+\rangle$ state,

$$DP = \frac{R^{(2)}_{\Psi^+} + R^{(1)}_{\Psi^+}}{R^{(2)}_{\Phi^\pm} + R^{(1)}_{\Phi^\pm,\Psi^-}} \tag{9}$$

where the detector's dark count is not included due to both negligible thermal energy at temperatures low enough for superconductivity compared to the semiconductor bandgap, and the lack of holes in the n-type region. The detection purity gives a good assessment of the detector's ability to distinguish between the desired $|\Psi^+\rangle$ Bell state detection, and carrier generation by the parasitic absorption of other Bell states and single photons. For photon energy that gives $\Omega^{HH} = \Delta$ and $\Delta\omega_p = 10meV$, $R^{(2)}_{\Phi^\pm}$ vanishes, and $R^{(2)}_{\Psi^+}$ attains a resonance at $\Delta\omega_{q_\mu,q_\nu} = 2\Delta(T)$, which splits the DP into two parts (Fig 3), for $\Delta\omega_{q_\mu,q_\nu} < 2\Delta(T)$ the detection purity is very high, due to the fact that neither one of the two photons has enough energy to reach the upper quasiparticle band, causing the one-photon rate to nearly vanish. On the other hand, for $\Delta\omega_{q_\mu,q_\nu} > 2\Delta(T)$ a high one photon detection rate severely deteriorates the entangled photon detection purity. In order to complete the picture, we assess the absorption coefficient $\alpha$ of the Bell-state detector using characteristic values of III-V semiconductors, and $\Delta\omega_{q_\mu,q_\nu} < 2\Delta(T)$ where the detection purity is high.

$$\alpha = \frac{256 S m_p^{HH} |B_{q_\mu} B_{q_\nu}|^2 \Delta^2}{v_g \left(\Delta\omega_{q_\mu,q_\nu} + \Omega^{HH}\right)^2 \left(\Delta\omega_{q_\mu,q_\nu} - \Omega^{HH}\right)^2} \tag{10}$$

where $v_g \approx c/3$ is the group velocity and $S \approx 10^{-8} cm^2$ is the contact surface between the superconductor and the PN junction. Assuming $\Omega^{HH} = 2\Delta\omega_{q_\mu,q_\nu} \approx 2\Delta$ we find the absorption coefficient to be similar to that of regular APDs with $\alpha \approx 10000\ cm^{-1}$.

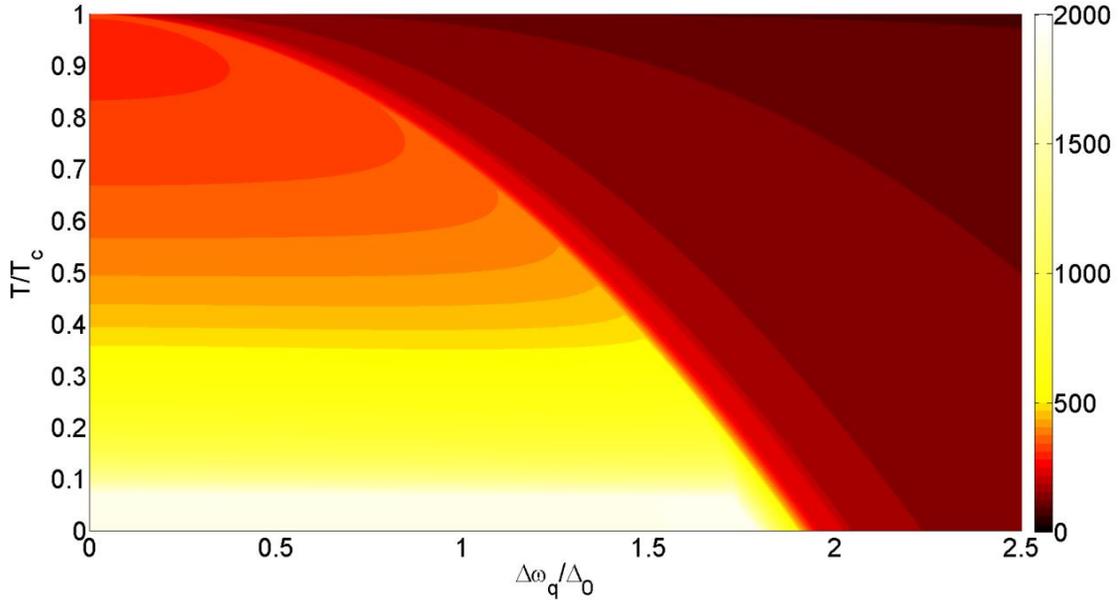

**Figure 3.** Detection purity $20\log(DP)$ dependence on normalized photon detuning energy $\Delta\omega_q/\Delta_0$ vs. normalized temperature $T/T_c$, with $\Delta_0 \equiv \Delta(T=0)$.

One of the most important parameters affecting DP is the HH-LH energy splitting $\Delta\omega_p$. No splitting of the hole energy bands, will drastically increase the false detection rate. Therefore, it is important to examine the detection purity dependence on the HH-LH energy splitting $\Delta\omega_p$ (Fig. 4). Keeping the total photon energy fixed and taking $\Omega^{HH} = \Delta$, small enough splitting energies such that $R^{(2)}_{\Phi^\pm}$ does not vanish, yield two minima which result from the two resonances of $R^{(2)}_{\Phi^\pm}$. Both minima correspond to virtual energy level coalescence with real energy levels, one for the HH level and one for the LH level. As the splitting grows the minima move to higher photon detuning energies. On the other hand the detection purity attains two peaks for the same reasons but for $R^{(2)}_{\Psi^+}$. One of these peaks vanishes when the LH part of $R^{(2)}_{\Psi^+}$ vanishes since the photon energy is too low. Once the LH-HH splitting is large enough (several $meV$), $R^{(2)}_{\Phi^\pm}$ vanishes, resulting in a high detection purity for a large range of photon detuning energies $\Delta\omega_q$, and especially high for small detuning energies. In practical devices the LH-HH splitting can reach tens of $meV$ depending on the QW thickness [27], therefore the use of such a device as a full Bell-state analyzer is very feasible.

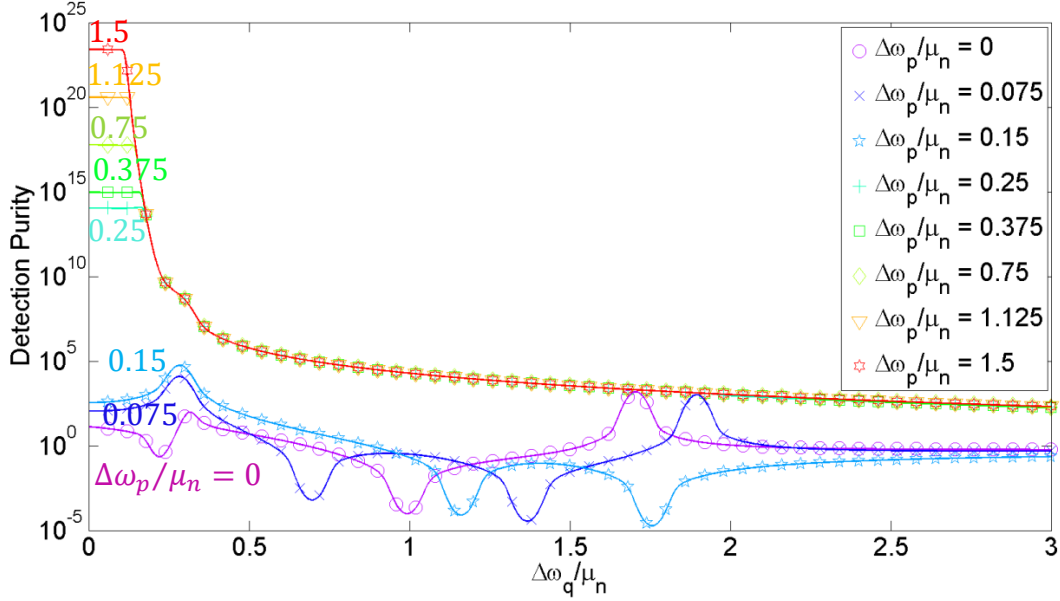

**Figure 4.** Detection purity dependence on normalized HH-LH splitting energy $\Delta\omega_p/\mu_n$ vs. normalized photon detuning energy $\Delta\omega_q/\mu_n$, with $\mu_n = 10 meV$.

Examining further the dependence of LH-HH energy splitting now with dependence on the photon bandwidth (BW) (Fig 5), for a wide-bandwidth pulse relative to the LH-HH energy splitting the detection purity is low since both rates $R^{(2)}_{\Psi^+}$ and $R^{(2)}_{\Phi^\pm}$ give a similar contribution. On the other hand, considering narrow-bandwidth photons relative to the LH-HH energy splitting ($BW/\Delta\omega_p < 0.5$) the detection purity is very high.

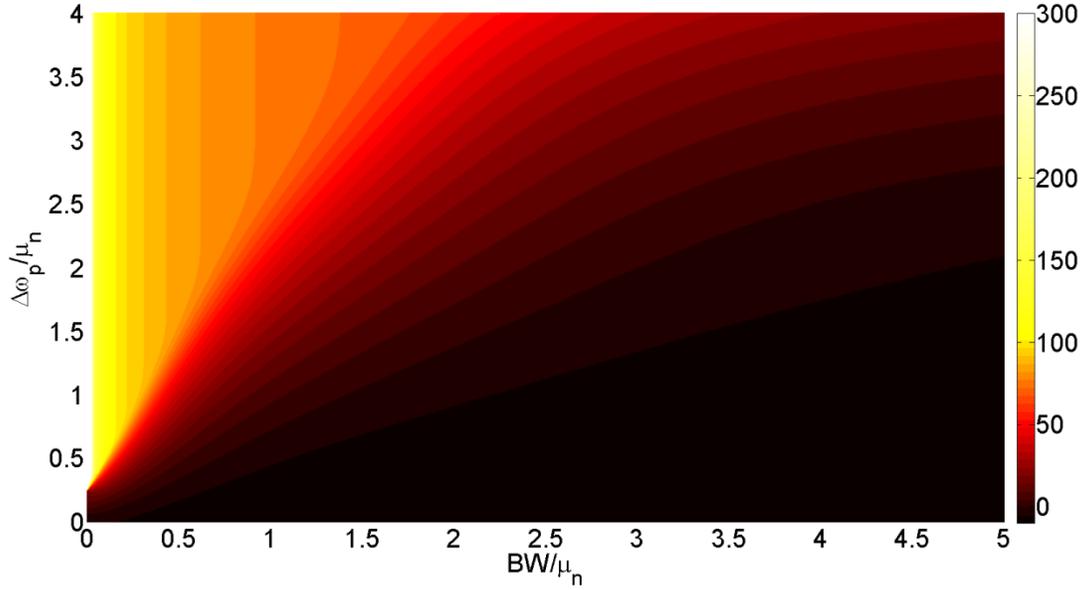

**Figure 5.** Detection purity dependence on normalized HH-LH splitting energy $\Delta\omega_p/\mu_n$ vs. normalized photon pulse energy bandwidth (using full width at half maximum) $BW/\mu_n$, with $\mu_n = 10 meV$.

The absence of single-electron transitions into energy levels inside the superconducting gap, has been demonstrated experimentally in numerous electrical tunneling measurements [23, 28] as well as in optical absorption experiments [29]. Whereas transitions into the superconducting gap are allowed only for Cooper pairs, e.g. in processes such as Andreev reflection [30]. However, various disorder types can introduce energy level broadening to the edges of the superconducting gap, and that in turn affects the two-photon and one-photon absorption ratio. In our model, we have taken two types of disorder into account. The first type is long-range disorder which accounts for slowly varying changes in the QW's potential and whose distribution is usually assumed to be Gaussian. Such disorder results in a Gaussian shaped broadening [31] of the one-photon absorption spectrum. The second type is short-range disorder which accounts for rapid spatial variation in the QW's potential. The effects of short-range disorder on the resulting one-photon absorption spectrum has been shown theoretically [32] and experimentally [33,34] to cause an exponential tail like broadening on the high-energy side of the spectrum. Combining both types of disorder yields a skewed-Gaussian like broadening which is Gaussian-like on the low-energy side of the spectrum and exponential-like on the high-energy side of the spectrum. Experimental results [33,34] have shown that for small ranges of energy, on the order of a few meV, the exponential tail is almost constant while also being ~3-4 orders of magnitude below the peak of the Gaussian broadening. Since the superconducting gap of low-$T_c$ superconductors is also on the order of a few meV, disorder-induced one-photon absorption in the superconducting gap is essentially energy-independent.

In order to emphasize the practical feasibility of our device, we have used the properties of InGaAs-GaAs QW as well as Nb or NbN as the superconductor. At 0K Nb has a superconducting gap of $\Delta_0 \equiv \Delta(T = 0)$ ~3.6 meV and a critical temperature of up to 9.25K [35], and NbN has a superconducting gap $\Delta_0$ of 5.2 meV and $T_c$ of 16K [36]. Modern fabrication techniques offer precision control over the thickness of the QW to a single molecular layer which yields small long-range disorder resulting in very narrow linewidths on the order of $\Delta E$ ~ 0.5meV [31]. This is an order of magnitude smaller than typical low-$T_c$ superconductor gap such as NbN, and two orders of magnitude smaller than those of high-$T_c$ superconductors. Whereas the uniformity of superconducting films has been demonstrated in density of states measurements showing narrowband features on the scale of less than 0.5meV [22]. Our calculations show (Fig. 6) that the exponential tail in the density of states due to short-range disorder contributes to parasitic one-photon absorption in the superconducting gap, which hardly depends on photon energy and is about 5 orders of magnitude weaker than the entangled photon pair detection.

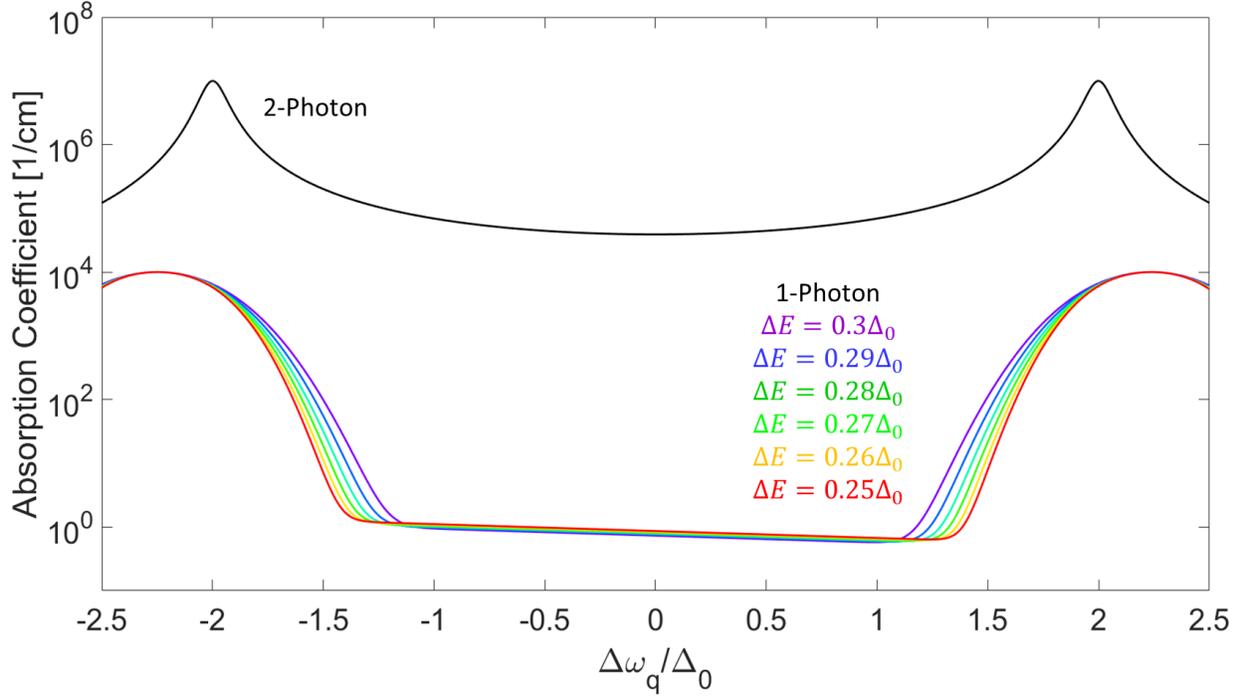

**Figure 6.** Calculated spectrum of the two-photon absorption (solid black line), and the disorder-induced one-photon absorption for different values of disorder broadening $\Delta E$ (colored lines) vs. normalized photon detuning energy.

This very small disorder-induced broadening also enables strongly-coupled light-matter interaction in semiconductor microcavities [37,38]. Therefore, for practically available QWs, the disorder-induced parasitic one-photon absorption is weaker than two-photon detection by at least five orders of magnitude.

In conclusion, our theoretical analysis shows that the proposed semiconductor-superconductor device has significant potential as a complete Bell-state analyzer. Due to the lifted degeneracy of the valance bands in QWs, Cooper-pair generation through entangled-photon absorption results in enhanced hole generation rate leading to high detection purity of the specified Bell state. The theoretically demonstrated Bell-state analyzer is shown to have high detection purity with very low false detection rates for a broad range of photon energies, enabling potential practical implementations of sophisticated quantum information applications.

# Photon Bell-state analysis based on semiconductor-superconductor structures

## Supplementary Material

### NUMBER OPERATOR EXPECTAION VALUE DERIVATION

Here we show a detailed derivation of the number operator expectation value to first and second order in perturbation theory leading to Eq. (3) and (4) in the paper, similar to a part of the derivation of the quantum Fokker-Planck equation [1]. We first start from the time evolution of an eigenstate of $H_0$ $|\chi_0\rangle$ with an interaction Hamiltonian $H_I$:

$$|\chi_t\rangle = \exp\left(-i \int_{-\infty}^{t} H_I(t') dt'\right) |\chi_0\rangle \tag{1}$$

Next, we use the time evolution of $|\chi_0\rangle$ to develop the change of a general number operator $N_i$ expectation value due to time evolution $\Delta\langle N_i\rangle = \langle \chi_t|N_i|\chi_t\rangle - \langle \chi_0|N_i|\chi_0\rangle$ given by:

$$\Delta\langle N_i\rangle = \left\langle \chi_0 \left| \exp\left(i \int_{-\infty}^{t} H_I(t') dt'\right) \left[N_i, \exp\left(-i \int_{-\infty}^{t} H_I(t') dt'\right)\right] \right| \chi_0 \right\rangle \tag{2}$$

Using the exponent's expansion to fourth order, coalescing quadratic terms in $H_I$ to $\langle N_i(1)\rangle$ and fourth order terms in $H_I$ to $\langle N_i(2)\rangle$, and using the fact that $\langle \chi_0|[N_i, A]|\chi_0\rangle$ vanishes for any operator $A$, if $|\chi_0\rangle$ is a Fock state, we get both Eq. (3) and (4) in the paper in their general form:

$$\langle N_i(1)\rangle = \int_{-\infty}^{t} dt_1 \int_{-\infty}^{t} dt_2 \langle \chi_0|H_I(t_1)[N_i, H_I(t_2)]|\chi_0\rangle \tag{3}$$

$$\begin{aligned}\langle N_i(2)\rangle = &\int_{-\infty}^{t_2} dt_1 \int_{-\infty}^{t} dt_2 \int_{-\infty}^{t} dt_3 \int_{-\infty}^{t_3} dt_4 \langle \chi_0|H_I(t_1)H_I(t_2)[N_i, H_I(t_3)H_I(t_4)]|\chi_0\rangle \\ &- \int_{-\infty}^{t_2} dt_1 \int_{-\infty}^{t_3} dt_2 \int_{-\infty}^{t} dt_3 \int_{-\infty}^{t} dt_4 \left\langle \chi_0 \left| \begin{matrix} H_I(t_1)H_I(t_2)H_I(t_3)[N_i, H_I(t_4)] \\ +H_I(t_4)[N_i, H_I(t_3)H_I(t_2)H_I(t_1)] \end{matrix} \right| \chi_0 \right\rangle \end{aligned} \tag{4}$$

# FIRST ORDER CALCULATION

Here we describe a detailed calculation of the first-order rate leading to Eq. (6) in the paper. We will first find the expectation value for $|\chi_0\rangle$ appearing in Eq. (3) in the paper, with $|Ph\rangle = \left|\sigma_{\mu_{q_\mu}}\sigma_{\nu_{q_\nu}}\right\rangle$, $q_\mu \neq q_\nu$. Since only one photon is absorbed the entanglement of the photons is irrelevant. Dividing the expectation value to two parts using the definition:

$$H_I = \sum_{k,q,\sigma,J}\left(B_{k,q}b_{-(k-q),-J}c_{k,J+\sigma}a^\dagger_{q,\sigma} + h.c.\right) = h_I + h_I^\dagger \tag{5}$$

We may refer to $\langle N_h^a(1)\rangle_{\chi_0} = \langle \chi_0|h_I(t_1)[N_h, h_I^\dagger(t_2)]|\chi_0\rangle$ as the photon absorption part and $\langle N_h^e(1)\rangle_{\chi_0} = \langle \chi_0|h_I^\dagger(t_1)[N_h, h_I(t_2)]|\chi_0\rangle$ as the photon emission part, where the rest of $h_I$ and $h_I^\dagger$ combinations vanish. Starting with the photon absorption part we get:

$$\langle N_h^a(1)\rangle_{\chi_0} = \sum_{k_1,k_2,q_1,q_2,J_1,J_2,\sigma_1,\sigma_2} B_{k_1,q_1}B^*_{k_2,q_2}e^{i\Omega_1 t_1 - i\Omega_2 t_2}I^{(1)}_{Ph}I^{(1)}_{FS}I^{(1)}_{BCS} \tag{6}$$

$$I^{(1)}_{Ph} = \langle Ph|a^\dagger_{q_1,\sigma_1}a_{q_2,\sigma_2}|Ph\rangle = \begin{pmatrix}\delta_{q_\mu,q_1,q_2}\delta_{\sigma_\mu,\sigma_1,\sigma_2} + \delta_{q_\nu,q_1,q_2}\delta_{\sigma_\nu,\sigma_1,\sigma_2}\\ +2\delta_{q_\mu,q_\nu,q_1,q_2}\delta_{\sigma_\mu,\sigma_\nu,\sigma_1,\sigma_2}\end{pmatrix} \tag{7}$$

$$I^{(1)}_{FS} = \sum_{p,J}\left\langle FS\left|b_{-(k_1-q_1),-J_1}\left[N_{p,J}, b^\dagger_{-(k_2-q_2),-J_2}\right]\right|FS\right\rangle = \delta_{k_1-q_1,k_2-q_2}\delta_{J_1,J_2}\left(1 - f^p_{k_1-q_1,J_1}\right) \tag{8}$$

$$I^{(1)}_{BCS} = \left\langle BCS\left|\begin{array}{l}\left(u_{k_1}e^{-iE_{k_1}t_1}\gamma_{k_1,J_1} - s_{J_1}v_{k_1}e^{iE_{k_1}t_1}\gamma^\dagger_{-k_1,\bar{J}_1}\right)\\ \times\left(u_{k_2}e^{iE_{k_2}t_2}\gamma^\dagger_{k_2,J_2} - s_{J_2}v_{k_2}e^{-iE_{k_2}t_2}\gamma_{-k_2,\bar{J}_2}\right)\end{array}\right|BCS\right\rangle$$

$$= \delta_{k_1,k_2}\delta_{J_1,J_2}\left[u^2_{k_1}e^{-iE_{k_1}(t_1-t_2)}\left(1 - f^n_{k_1}\right) + v^2_{k_1}e^{iE_{k_1}(t_1-t_2)}f^n_{k_1}\right] \tag{9}$$

where $\Omega_i = \omega_{q_i} - \epsilon_{k_i-q_i,J_i} - \tilde{\mu}_n = \omega_{q_i} - \tilde{\mu}_n - \tilde{\mu}_p - \xi_p(k_i - q_i, J_i)$ with $\tilde{\mu}_n = E_c + \mu_n$, $\tilde{\mu}_p = E_v + \mu_p$, $\epsilon_{p,J} = \frac{p^2}{2m_{p,J}} + E_v + \Delta\omega_{p,J}$ and $\xi_p(k,J) = \frac{k^2}{2m_{k,J}} - \mu_p + \Delta\omega_{p,J}$, $f^n_k = \left(e^{-\beta E_k} + 1\right)^{-1}$ and

$f_{k,J}^p = \left(e^{-\beta \xi_p(k,J)} + 1\right)^{-1}$ are the Bogoliubov quasiparticles and holes Fermi-Dirac distribution, respectively. Substituting Eqs. (7), (8), and (9) into Eq. (6), we obtain:

$$\langle N_h^a(1) \rangle_{\chi_0} = \sum_{k,q,J} |B_{k,q}|^2 e^{i\Omega(t_1 - t_2)} \left(\delta_{q_\mu,q} + \delta_{q_\nu,q}\right) \left(1 - f_{k-q,J}^p\right) \begin{bmatrix} u_k^2 e^{-iE_k(t_1 - t_2)}(1 - f_k^n) \\ +v_k^2 e^{iE_k(t_1 - t_2)} f_k^n \end{bmatrix} \quad (10)$$

Using the same process for the photon emission part and neglecting stimulated emission due to $\sum_{k,q} 1 \gg 1$ we obtain:

$$\langle N_h^e(1) \rangle_{\chi_0} = -2 \sum_{k,q,J} |B_{k,q}|^2 e^{-i\Omega(t_1 - t_2)} f_{k-q,J}^p \begin{bmatrix} u_k^2 e^{iE_k(t_1 - t_2)} f_k^n \\ +v_k^2 e^{-iE_k(t_1 - t_2)}(1 - f_k^n) \end{bmatrix} \quad (11)$$

Combining the results from Eqs. (10) and (11), carrying out the time integrations using the long time approximation and taking a time derivative we obtain the first order rate:

$$R^{(1)} = 2\pi \sum_{k,q,J,\sigma} |B_{k,q}|^2 \begin{bmatrix} \left(\delta_{q_\mu,q} + \delta_{q_\nu,q}\right)\left(1 - f_{k-q}^p\right) \begin{bmatrix} |u_k|^2 \delta(\Omega - E_k)(1 - f_k^n) \\ +|v_k|^2 \delta(\Omega + E_k) f_k^n \end{bmatrix} \\ -2f_{k-q}^p [|u_k|^2 \delta(\Omega - E_k) f_k^n + |v_k|^2 \delta(\Omega + E_k)(1 - f_k^n)] \end{bmatrix} \quad (12)$$

In order to obtain the final result we transform the summation over k to integration on 2D using $\sum_k \to \frac{S}{(2\pi)^2} \int d^2k = \frac{S}{2\pi} \int k dk$, with S being the contact surface between the superconductor and the PN junction, neglecting $B_{k,q}$ dependence on k and taking $k \gg q$ we get the final expression for the first order hole creation rate:

$$R^{(1)} = Sm_p \sum_{q,J,\xi_p} |B_q|^2 sign(\widetilde{\omega}_q + \mu_J)\Theta(\xi_p + \mu_p)\Theta\left(\overline{m}_J^2(\widetilde{\omega}_q + \mu_J)^2 + \Delta^2(1 - \overline{m}_J^2)\right)$$
$$\times \begin{bmatrix} \left(\delta_{q_\mu,q} + \delta_{q_\nu,q}\right)\left(1 - f_{\xi_p}^p\right)\left[(1 - f_{\xi_n}^n)\Theta(\widetilde{\omega}_q - \xi_p) - f_{\xi_n}^n \Theta(\xi_p - \widetilde{\omega}_q)\right] \\ -2f_{\xi_p}^p [f_{\xi_n}^n \Theta(\widetilde{\omega}_q - \xi_p) - (1 - f_{\xi_n}^n)\Theta(\xi_p - \widetilde{\omega}_q)] \end{bmatrix} \quad (13)$$

where $\xi_p^{(\pm)} = \frac{1}{1-\bar{m}_J^2}\left[\tilde{\omega}_q + \bar{m}_J^2 \mu_J \pm \sqrt{\bar{m}_J^2(\tilde{\omega}_q + \mu_J)^2 + \Delta^2(1-\bar{m}_J^2)}\right]$, $f_{\xi_p}^p = (e^{\beta\xi_p} + 1)^{-1}$ and $f_{\xi_n}^n = \left[\exp(\beta\sqrt{\xi_n^2 + \Delta^2}) + 1\right]^{-1}$, and we've defined $\bar{m}_J \equiv \frac{m_{p,J}}{m_n}$, $\mu_J \equiv \mu_p - \Delta\omega_{p,J} - \frac{m_n}{m_{p,J}}\mu_n$, $\tilde{\omega}_q \equiv \omega_q - (\tilde{\mu}_n + \tilde{\mu}_p)$. One may consider another very good approximation of taking $f_{\xi_p}^p \to 0$ since $\beta\xi_p \gg 1$ for the low temperatures demanded for superconductivity and get the following simplified expression:

$$R^{(1)} = Sm_p \sum_{J,\xi_p} |B_q|^2 \text{sign}(\tilde{\omega}_{q_\mu} + \mu_J) \Theta(\xi_p + \mu_p)\Theta\left(\bar{m}_J^2(\tilde{\omega}_{q_\mu} + \mu_J)^2 + \Delta^2(1-\bar{m}_J^2)\right) \\ \times \left[(1-f_{\xi_n}^n)\Theta(\tilde{\omega}_{q_\mu} - \xi_p) - f_{\xi_n}^n \Theta(\xi_p - \tilde{\omega}_{q_\mu})\right] + (q_\mu \to q_\nu) \quad (14)$$

Important to note that this result is true for all Bell states meaning $R^{(1)}_{\Phi^\pm,\Psi^\pm} = R^{(1)}$ since it's a one photon process.

## SECOND-ORDER CALCULATION

We now detail the calculation of the second-order rate leading to Eqs. (7) and (8) in the paper. Using a similar method of calculation as for the first order rate starting with $|Ph\rangle = |\Psi^\pm\rangle$. We again use Eq. (5) to divide the expectation value $\langle\chi_0|H_I(t_1)H_I(t_2)[N_h, H_I(t_3)H_I(t_4)]|\chi_0\rangle$ to two parts, referring to $\langle N_{h,a}^{\Psi^\pm}(2)\rangle_{\chi_0} = \langle\chi_0|h_I(t_1)h_I(t_2)[N_h, h_I^\dagger(t_3)h_I^\dagger(t_4)]|\chi_0\rangle$ as the photon absorption part and $\langle N_{h,e}^{\Psi^\pm}(2)\rangle_{\chi_0} = \langle\chi_0|h_I^\dagger(t_1)h_I^\dagger(t_2)[N_h, h_I(t_3)h_I(t_4)]|\chi_0\rangle$ as the photon emission part, where again the rest of $h_I$ and $h_I^\dagger$ combinations vanish. As for the second integral term in Eq. (4) in the paper, we will refer to it in a later stage of our calculation. Starting with the photon absorption part we get:

$$\left\langle N_{h,a}^{\Psi^\pm}(2) \right\rangle_{\chi_0} = \sum_{\substack{k_1,\ldots,k_4,q_1,\ldots,q_4 \\ J_1,\ldots,J_4,\sigma_1,\ldots,\sigma_4}} e^{i\Omega_1 t_1 + i\Omega_2 t_2 - i\Omega_3 t_3 - i\Omega_4 t_4} B_{k_1,q_1} B_{k_2,q_2} B^*_{k_3,q_3} B^*_{k_4,q_4} I_{Ph}^{(2)} I_{FS}^{(2)} I_{BCS}^{(2)} \quad (15)$$

$$I_{Ph}^{(2)} = \frac{1}{2}\left[\begin{pmatrix}\delta_{q_1,q_\mu}\delta_{\sigma_1,R}\delta_{q_2,q_\nu}\delta_{\sigma_2,L} \\ \pm \delta_{q_1,q_\mu}\delta_{\sigma_1,L}\delta_{q_2,q_\nu}\delta_{\sigma_2,R}\end{pmatrix} \pm (\mu \leftrightarrow \nu)\right]\left[\begin{pmatrix}\delta_{q_4,q_\mu}\delta_{\sigma_4,R}\delta_{q_3,q_\nu}\delta_{\sigma_3,L} \\ \pm \delta_{q_4,q_\mu}\delta_{\sigma_4,L}\delta_{q_3,q_\nu}\delta_{\sigma_3,R}\end{pmatrix} \pm (\mu \leftrightarrow \nu)\right] \quad (16)$$

$$I_{FS}^{(2)} = 2\left(1 - f^p_{k_1-q_1,J_1}\right)\left(1 - f^p_{k_2-q_2,J_2}\right)\begin{pmatrix}\delta_{k_1-q_1,k_4-q_4}\delta_{J_1,J_4}\delta_{k_2-q_2,k_3-q_3}\delta_{J_2,J_3} \\ -\delta_{k_1-q_1,k_3-q_3}\delta_{J_1,J_3}\delta_{k_2-q_2,k_4-q_4}\delta_{J_2,J_4}\end{pmatrix} \quad (17)$$

$$I_{BCS}^{(2)} \approx u_{k_1}v_{k_1}u_{k_3}v_{k_3}\delta_{J_1+\sigma_1,\overline{J_2+\sigma_2}}\delta_{J_3+\sigma_3,\overline{J_4+\sigma_4}}\delta_{k_1,-k_2}\delta_{k_3,-k_4}S_{J_1+\sigma_1}S_{J_3+\sigma_3}$$

$$\times \begin{bmatrix} e^{-iE_{k_1}(t_1-t_2)}e^{iE_{k_3}(t_3-t_4)}\left(1-f^n_{k_1}\right)f^n_{k_3} - e^{-iE_{k_1}(t_1-t_2)}e^{-iE_{k_3}(t_3-t_4)}\left(1-f^n_{k_1}\right)\left(1-f^n_{k_3}\right) \\ -e^{iE_{k_1}(t_1-t_2)}e^{iE_{k_3}(t_3-t_4)}f^n_{k_1}f^n_{k_3} + e^{iE_{k_1}(t_1-t_2)}e^{-iE_{k_3}(t_3-t_4)}f^n_{k_1}\left(1-f^n_{k_3}\right) \end{bmatrix} \quad (18)$$

where $I_{Ph}^{(2)}$, $I_{FS}^{(2)}$ and $I_{BCS}^{(2)}$ are defined similarly to their first order counterparts. $I_{BCS}^{(2)}$ is given by an approximated term since it only holds the most dominant contributions coming from Cooper pairs. Substituting Eqs. (16), (17), and (18) into Eq. (15), we obtain:

$$\left\langle N_{h,a}^{\Psi^\pm}(2) \right\rangle_{\chi_0} = \sum_{k} \left|B_{k,q_\mu}B_{k,q_\nu}\right|^2 (1 \pm 1)\left[\begin{matrix} e^{i\Omega_\mu^{LH}t_1+i\Omega_\nu^{LH}t_2-i\Omega_\nu^{LH}t_3-i\Omega_\mu^{LH}t_4} \\ +e^{i\Omega_\mu^{LH}t_1+i\Omega_\nu^{LH}t_2-i\Omega_\mu^{LH}t_3-i\Omega_\nu^{LH}t_4} \end{matrix} + (\mu \leftrightarrow \nu)\right]u_k^2 v_k^2$$

$$\times 2\left(1 - f^p_{k,LH}\right)^2 \begin{bmatrix} e^{iE_k(t_1-t_2+t_3-t_4)}(f^n_k)^2 + e^{-iE_k(t_1-t_2+t_3-t_4)}(1-f^n_k)^2 \\ -\left(e^{iE_k(t_1-t_2-t_3+t_4)} + e^{-iE_k(t_1-t_2-t_3+t_4)}\right)f^n_k(1-f^n_k) \end{bmatrix} + (LH \leftrightarrow HH) \quad (19)$$

where $\Omega_i^{HH(LH)} = \omega_{q_i} - \epsilon_{k-q_i,J=\pm\frac{3}{2}\left(\frac{1}{2}\right)} - \tilde{\mu}_n = \omega_{q_i} - \tilde{\mu}_n - \tilde{\mu}_p - \xi_p\left(k - q_i, J = \pm\frac{3}{2}\left(\frac{1}{2}\right)\right)$ and $f^p_{k,HH(LH)} = f^p_{k,J=\pm\frac{3}{2}\left(\frac{1}{2}\right)}$. Given this result we can immediately see that $\left\langle N_{h,a}^{\Psi^-}(2) \right\rangle_{\chi_0} = 0$ and only $\left\langle N_{h,a}^{\Psi^+}(2) \right\rangle_{\chi_0}$ gives a finite result. Using the same derivation for the photon emission part and neglecting stimulated emission we get:

$$\left\langle N_{h,e}^{\Psi^\pm}(2) \right\rangle_{\chi_0} = -\sum_{k,q_1,q_2} \left|B_{k,q_1}B_{k,q_2}\right|^2 \begin{bmatrix} 4f^p_{k,LH}e^{-i\Omega_1^{LH}t_1-i\Omega_2^{LH}t_2+i\Omega_1^{LH}t_3+i\Omega_2^{LH}t_4} \\ 8f^p_{k,HH}e^{-i\Omega_1^{LH}t_1-i\Omega_2^{HH}t_2+i\Omega_1^{LH}t_3+i\Omega_2^{HH}t_4} \end{bmatrix} + (t_3 \leftrightarrow t_4) \quad (20)$$

$$\times f_{k,LH}^p u_k^2 v_k^2 \begin{bmatrix} e^{iE_k(t_1-t_2+t_3-t_4)}(f_k^n)^2 + e^{-iE_k(t_1-t_2+t_3-t_4)}(1-f_k^n)^2 \\ -\left(e^{iE_k(t_1-t_2-t_3+t_4)} + e^{-iE_k(t_1-t_2-t_3+t_4)}\right)f_k^n(1-f_k^n) \end{bmatrix} + (LH \leftrightarrow HH)$$

Before we continue the derivation for the first integral in Eq. (4) in the paper, we now return to treat the second integral term noted before. Pairing all non-vanishing $h_I$ and $h_I^\dagger$ combinations:

$$\langle \chi_0 | H_I(t_1)H_I(t_2)H_I(t_3)[N_i, H_I(t_4)] + H_I(t_4)[N_i, H_I(t_3)H_I(t_2)H_I(t_1)] | \chi_0 \rangle =$$

$$\left\langle \chi_0 \left| \begin{matrix} h_I^\dagger(t_1)h_I(t_2)h_I(t_3)[N_h, h_I^\dagger(t_4)] + h_I^\dagger(t_4)[N_h, h_I(t_3)h_I(t_2)h_I^\dagger(t_1)] \\ +h_I(t_1)h_I^\dagger(t_2)h_I(t_3)[N_h, h_I^\dagger(t_4)] + h_I^\dagger(t_4)[N_h, h_I(t_3)h_I^\dagger(t_2)h_I(t_1)] \\ +h_I(t_1)h_I(t_2)h_I^\dagger(t_3)[N_h, h_I^\dagger(t_4)] + h_I^\dagger(t_4)[N_h, h_I^\dagger(t_3)h_I(t_2)h_I(t_1)] \end{matrix} \right| \chi_0 \right\rangle \quad (21)$$

$$+ \left\langle \chi_0 \left| \begin{matrix} h_I(t_1)h_I^\dagger(t_2)h_I^\dagger(t_3)[N_h, h_I(t_4)] + h_I(t_4)[N_h, h_I^\dagger(t_3)h_I^\dagger(t_2)h_I(t_1)] \\ +h_I^\dagger(t_1)h_I(t_2)h_I^\dagger(t_3)[N_h, h_I(t_4)] + h_I(t_4)[N_h, h_I^\dagger(t_3)h_I(t_2)h_I^\dagger(t_1)] \\ +h_I^\dagger(t_1)h_I^\dagger(t_2)h_I(t_3)[N_h, h_I(t_4)] + h_I(t_4)[N_h, h_I(t_3)h_I^\dagger(t_2)h_I^\dagger(t_1)] \end{matrix} \right| \chi_0 \right\rangle$$

Using the relations $[N_h, b_{\boldsymbol{p},J}] = -b_{\boldsymbol{p},J}$ and $[N_h, b_{\boldsymbol{p},J}^\dagger] = b_{\boldsymbol{p},J}^\dagger$ we get the following result:

$$\langle \chi_0 | H_I(t_1)H_I(t_2)H_I(t_3)[N_i, H_I(t_4)] + H_I(t_4)[N_i, H_I(t_3)H_I(t_2)H_I(t_1)] | \chi_0 \rangle =$$

$$\left\langle \chi_0 \left| \begin{matrix} h_I^\dagger(t_1)h_I(t_2)h_I(t_3)h_I^\dagger(t_4) - h_I^\dagger(t_4)h_I(t_3)h_I(t_2)h_I^\dagger(t_1) \\ +h_I(t_1)h_I^\dagger(t_2)h_I(t_3)h_I^\dagger(t_4) - h_I^\dagger(t_4)h_I(t_3)h_I^\dagger(t_2)h_I(t_1) \\ +h_I(t_1)h_I(t_2)h_I^\dagger(t_3)h_I^\dagger(t_4) - h_I^\dagger(t_4)h_I^\dagger(t_3)h_I(t_2)h_I(t_1) \end{matrix} \right| \chi_0 \right\rangle \quad (22)$$

$$- \left\langle \chi_0 \left| \begin{matrix} h_I(t_1)h_I^\dagger(t_2)h_I^\dagger(t_3)h_I(t_4) - h_I(t_4)h_I^\dagger(t_3)h_I^\dagger(t_2)h_I(t_1) \\ +h_I^\dagger(t_1)h_I(t_2)h_I^\dagger(t_3)h_I(t_4) - h_I(t_4)h_I^\dagger(t_3)h_I(t_2)h_I^\dagger(t_1) \\ +h_I^\dagger(t_1)h_I^\dagger(t_2)h_I(t_3)h_I(t_4) - h_I(t_4)h_I(t_3)h_I^\dagger(t_2)h_I^\dagger(t_1) \end{matrix} \right| \chi_0 \right\rangle$$

Each pair cancels one another's Cooper pair dependent terms, therefore we only get a negligible contribution from this integral that doesn't include Cooper-pairs, thus it can be ignored.

Going back to the derivation of the first integral term in Eq. (4) in the paper, we can now carry out the time integrations using the long time approximation as before to the results from Eqs. (19) and (20), and taking a time derivative we can obtain their contribution to $R_{\Psi^+}^{(2)}$:

$$R_{\Psi^+}^{a(2)} = 32\pi \sum_k \left|B_{k,q_\mu}B_{k,q_\nu}\right|^2 \delta\left(\Omega_\mu^{LH} + \Omega_\nu^{LH}\right) \frac{\left(1 - f_{k,LH}^p\right)^2 \Delta^2}{\left(\Omega_\mu^{LH} + E_k\right)^2 \left(\Omega_\mu^{LH} - E_k\right)^2} + (LH \leftrightarrow HH) \quad (23)$$

$$R_{\Psi^\pm}^{e(2)} = -8\pi \sum_{k,q_1,q_2} \left|B_{k,q_1} B_{k,q_2} \frac{\Delta}{E_k}\right|^2 f_k^{p(LH)} \begin{bmatrix} \delta(\Omega_1^{LH} + \Omega_2^{LH}) f_k^{p(LH)} \\ +2\delta(\Omega_1^{LH} + \Omega_2^{HH}) f_k^{p(HH)} \end{bmatrix}$$

$$\times \left[ \frac{f_k^n}{(\Omega_1^{LH} - E_k)^2} - \frac{1}{(\Omega_1^{LH} + E_k)(\Omega_1^{LH} - E_k)} + \frac{(1 - f_k^n)}{(\Omega_1^{LH} + E_k)^2} \right] + (LH \leftrightarrow HH) \quad (24)$$

Following the same steps as for the first order derivation, we transform the summation over k to integration, neglecting $B_{k,q}$ dependence on k and taking $k \gg q$ we get:

$$R_{\Psi^+}^{a(2)} = 256 \left|B_{q_\mu} B_{q_\nu}\right|^2 Sm_p^{LH} \frac{\left(1 - f_{\xi_p}^p\right)^2 \Delta^2 \Theta\left(\omega_{q_\mu} + \omega_{q_\nu} - 2(E_g + \Delta\omega_p^{LH} + \mu_n)\right)}{\left(\Delta\omega_{q_\mu,q_\nu} + \Omega^{LH}\right)^2 \left(\Delta\omega_{q_\mu,q_\nu} - \Omega^{LH}\right)^2} \quad (25)$$

$$+ (LH \leftrightarrow HH)$$

where $\xi_p = \frac{1}{2}\left(\omega_{q_i} + \omega_{q_j} - 2\tilde{\mu}_n - 2\tilde{\mu}_p\right)$, $\left(\Omega^{LH(HH)}\right)^2 = \left[\frac{m_p^{LH(HH)}}{m_n}\left(\omega_{q_\mu} + \omega_{q_\nu} - 2(E_g + \Delta\omega_p^{LH(HH)} + \mu_n)\right) - 2\mu_n\right]^2 + 4\Delta^2$, $\Delta\omega_{q_\mu,q_\nu} = \omega_{q_\mu} - \omega_{q_\nu}$ and $f_{\xi_p}^p = \left(e^{\beta\xi_p} + 1\right)^{-1}$. As for the emission term we neglect it, since we can take the holes Fermi-Dirac distribution to zero to a good approximation, and by using this approximation on the absorption term rate we get:

$$R_{\Psi^+}^{(2)} = 256 Sm_p^{LH} \frac{\left|B_{q_\mu} B_{q_\nu}\right|^2 \Delta^2 \Theta\left(\omega_{q_\mu} + \omega_{q_\nu} - 2(E_g + \Delta\omega_p^{LH} + \mu_n)\right)}{\left(\Delta\omega_{q_\mu,q_\nu} + \Omega^{LH}\right)^2 \left(\Delta\omega_{q_\mu,q_\nu} - \Omega^{LH}\right)^2} + (LH \leftrightarrow HH) \quad (26)$$

Using the same derivation for $|Ph\rangle = |\Phi^{\pm}\rangle$, and referring to $\left\langle N_{h,a}^{\Phi^{\pm}}(2)\right\rangle_{\chi_0} = \langle\chi_0|h_I(t_1)h_I(t_2)[N_h, h_I^{\dagger}(t_3)h_I^{\dagger}(t_4)]|\chi_0\rangle$ as the photon absorption part and $\left\langle N_{h,e}^{\Phi^{\pm}}(2)\right\rangle_{\chi_0} = \langle\chi_0|h_I^{\dagger}(t_1)h_I^{\dagger}(t_2)[N_h, h_I(t_3)h_I(t_4)]|\chi_0\rangle$ as the photon emission part, we obtain for the absorption term:

$$\left\langle N_{h,a}^{\Phi^{\pm}}(2)\right\rangle_{\chi_0} = 4\sum_k |B_{k,q_\mu}B_{k,q_\nu}|^2 \left[\begin{matrix}e^{i\Omega_\mu^{LH}t_1 + i\Omega_\nu^{HH}t_2 - i\Omega_\nu^{HH}t_3 - i\Omega_\mu^{LH}t_4}\\ +e^{i\Omega_\mu^{LH}t_1 + i\Omega_\nu^{HH}t_2 - i\Omega_\mu^{LH}t_3 - i\Omega_\nu^{HH}t_4}\end{matrix} + (\mu\leftrightarrow\nu)\right] u_k^2 v_k^2$$

$$\times (1-f_{k,LH}^p)(1-f_{k,HH}^p)\begin{bmatrix}e^{iE_k(t_1-t_2+t_3-t_4)}(f_k^n)^2 + e^{-iE_k(t_1-t_2+t_3-t_4)}(1-f_k^n)^2\\ -\left(e^{iE_k(t_1-t_2-t_3+t_4)} + e^{-iE_k(t_1-t_2-t_3+t_4)}\right)f_k^n(1-f_k^n)\end{bmatrix} \quad (27)$$

$$+ (LH \leftrightarrow HH)$$

This expression is very similar to the expression for $\left\langle N_{h,a}^{\Psi^{\pm}}(2)\right\rangle_{\chi_0}$; however, here both + and – terms are finite. The expression for the emission term remains the same as before, since spontaneous emission is not affected by the state $|Ph\rangle$. In a very similar manner we get to the final result for the rate $R_{\Phi^{\pm}}^{(2)}$,

$$R_{\Phi^{\pm}}^{(2)} = 256Sm_p \frac{|B_{q_\mu}B_{q_\nu}|^2 \Delta^2\Theta\left(\omega_{q_\mu} + \omega_{q_\nu} - 2\left(E_g + \frac{1}{2}\Delta\omega_p + \mu_n\right)\right)}{\left(\Delta\omega_{q_\mu,q_\nu}^{LH} + \Omega\right)^2 \left(\Delta\omega_{q_\mu,q_\nu}^{LH} - \Omega\right)^2} + (LH \leftrightarrow HH) \quad (28)$$